# Jeeva: Enterprise Grid Enabled Web Portal for Protein Secondary Structure Prediction

Chao Jin [#], Jayavardhana Gubbi [*], Rajkumar Buyya [#], Marimuthu Palaniswami [*]

[#] Grid Computing and Distributed Systems (GRIDS) Laboratory
Department of Computer Science and Software Engineering
The University of Melbourne, Melbourne, VIC 3010, Australia
{chaojin, raj}@csse.unimelb.edu.au

[*] Department of Electrical and Electronic Engineering
The University of Melbourne, Melbourne, VIC 3010, Australia
{jrgl, swami}@ee.unimelb.edu.au

*Abstract*—This paper presents a Grid portal for protein secondary structure prediction developed by using services of Aneka, a .NET-based enterprise Grid technology. The portal is used by research scientists to discover new prediction structures in a parallel manner. An SVM (Support Vector Machine)-based prediction algorithm is used with 64 sample protein sequences as a case study to demonstrate the potential of enterprise Grids.

## I. INTRODUCTION

The structure of protein plays a key role in the structure-based design of drugs for the treatment of various diseases. However, it is still a challenge to find out protein structure based on its sequence, and the dependence on experimental methods may not yield protein structures fast enough to keep up with the requirement of current industry. Fortunately, the energy landscape theory [24] enables a framework for the development of algorithms to predict the structure of unknown proteins based on their sequence, which is known as protein structure prediction.

From the perspective of computer science, protein structure prediction is a computing intensive task [5]. Since the prediction of protein structure is a complex task, it is usually sub-divided into two phases. The first one is secondary structure prediction and the second one is super secondary structure prediction, leading to tertiary structure, i.e., the specific atomic positions in three-dimensional space. As the first phase of protein structure prediction, accurate secondary structure prediction is a key element for correctly acquiring tertiary structure.

A large number of algorithms [2][6][9][11] have been proposed for protein secondary structure prediction. To facilitate the collaboration between protein scientists across the world, it is a necessity for researchers to share their algorithms and results with colleagues dispersed at different geographical locations. Furthermore, to speed up the process of finding out new protein structures, we need a proper computational platform which simplifies the development of new prediction algorithms and improves the efficiency at the same time. For example, machine learning methods are currently used for secondary structure prediction. In particular, SVM (Support Vector Machines) based prediction has many advantages compared with other solutions [13]. However, its computing intensive nature demands an improvement on its efficiency by parallel processing.

In order to address the above two issues, Grid computing offers important solutions. Grid computing [10] provides faster computation facilities for minimizing the time required for solving problems, supporting on-demand access to distributed computing resources from multiple organisations, and enabling the creation of community computing application portal services.

This paper proposes and presents the design, development and deployment of an interactive web-based portal, called Jeeva, for quick discovery of protein secondary structure prediction. In particular, our platform aims to support the following capabilities:

- A collaborative environment to encourage and assist the deployment of new prediction algorithms in a parallel way, particularly for those amateur researchers with less well-developed skills and expertise on parallel programming.
- An easy for use environment for public users to access prediction algorithms released in our web portal and to manage their prediction history results in an online manner.

Jeeva web portal system consists of an interactive web interface and a Grid middleware. With the interactive web interface, users can submit prediction requests for protein secondary structures, collect results, and manage the history of prediction data. By means of the Grid middleware, researchers can not only deploy their prediction applications in a distributed environment easily, but also monitor and manage the execution in the distributed environment. The Grid enablement of Jeeva is achieved by using Aneka [27], which is a .NET-based Grid software system for the creation of enterprise Grid environments.

We use an SVM-based protein secondary structure prediction algorithm [13] as a case study to show the usage of Jeeva, and experiments to evaluate the performance and scalability of our platform.

The remainder of this paper is organized as follows. Section II provides a discussion on related work. Section III describes the background on SVM-based prediction. Section IV presents the architecture, design, and implementation of Jeeva. Section V shows the experimental evaluation of the system through the chosen SVM based prediction algorithm. Section VI concludes the paper with pointers to future work.

## II. RELATED WORK

Protein secondary structure prediction is based on the prediction of protein 1-D structure from the sequence of aminoacid residues in the target protein [3]. Several methods have been proposed to find out the secondary structure based on physico-chemical properties and homology. The most popular secondary structure prediction methods currently in use include [1], [7], [11], [16], [19]. A detailed review of secondary structure algorithms until the year 2000 can be found in [1].

Recently, some significant work has been done on secondary structure prediction using Support Vector Machines. Hua and Sun [22] used SVMs and profiles of the multiple alignments from HSSP database as features and reported a $Q_3$ score as 73.5% on the CB513 dataset [11]. In 2003, Ward [15] reported 77% with PSI-BLAST [21] profiles on a small set of proteins. In the same year Kim and Park [9] reported an accuracy of 76.6% on the CB513 dataset using PSI-BLAST Position Specific Scoring Matrix (PSSM). Nguyen and Rajapakse [17][18] explored several multi-class recognition schemes and reported a highest accuracy of 72.8% on RS126 dataset using a two stage SVM. Guo [14] used a dual layered SVM with profiles and reported a highest accuracy of 75.2% on the CB513 dataset. More recently, Hu [8] reported the highest accuracy of 78.8% on a RS126 dataset using a novel encoding scheme.

A few of the above methods are made available in web servers for online access and utilization. As far as the authors are aware, none of the secondary structure prediction systems based on SVM is available through the web service technology. A few other servers supporting homology modeling, neural networks and hidden markov models, include PHD [2], PROF-King [19], PSIPred [7], JPred [11], SAMT99-Sec [16], and SCRATCH [12]. The SCRATCH web server uses a SVM for disulphide bridge prediction and a recursive neural network for secondary structure prediction.

Predictor@Home [20] is using contributory resources for predicting the tertiary structure of proteins over the BOINC [6] platform. However, their secondary prediction algorithm runs locally in a sequential manner.

## III. BACKGROUND ON SVM-BASED PREDICTION

An SVM based secondary structure prediction algorithm is used in [13]. Briefly, this method investigates the effect of the physico-chemical and statistical properties on protein secondary structure prediction along with evolutionary information in the form of position specific scoring matrix (PSSM). SVMs [26] are usually employed for classification and the outputs of SVM are converted to posterior probabilities for multi-class classification. For the web enabled system, we use the Chou-Fasman parameters and physico-chemical parameters along with evolutionary information in the form of position specific scoring matrix (PSSM) as features. The SVM implementation used in Jeeva is SVMLight [25].

It is well known that testing new input data by using SVM is relatively slow compared to other machine learning approaches. In case of protein structure prediction, the problem becomes more complex as the training size of the data is very large, i.e. in the order of tens of thousands. For multi-class classification in secondary structure prediction, many SVMs are required. In our case, for three class classification, six SVM models are required. This considerably increases the computational complexity. As each of these classifiers is independent of each other, it is obvious that parallelizing them has profound effects in the final time taken for predicting the secondary structure. In our current web enabled system, each classifier is taken as an independent task supported by the task programming model in Aneka.

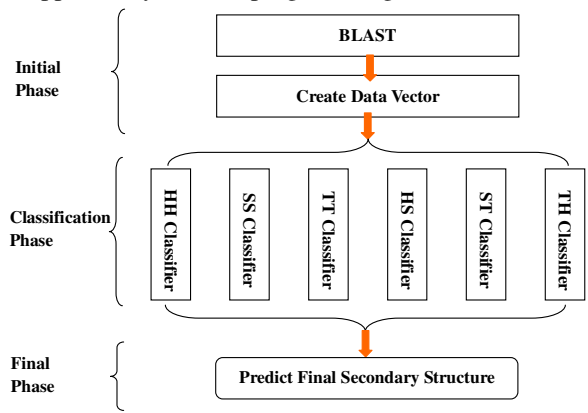

Fig. 1 Flow chart of The SVM based Prediction Algorithm.

**Fig. 1** illustrates the flow chart of the SVM based algorithm. There are 3 phases: initial, classification and final prediction phases. During the initial phase, the algorithm reads a protein sequence, submits it to PSI-BLAST [21] to obtain the PSSM features and finally generates feature vector for classification.

A new dataset from CATH [4] (version 2.6.0) is created. This set has been used to train the system for all predictions[1]. At the first stage of dataset preparation, proteins with sequence length greater than 40 and resolution of at least 2 Ang are selected. We use UniqueProt [23] with an HSSP-value of 0 to eliminate identical sequences. Out of 10,000 proteins, 504 proteins which have the sequence identity of less than 15% are retained. There are 97,593 residues with the average sequence length of 194.

---

[1] http://www.ee.unimelb.edu.au/ISSNIP/bioinf/

The classification phase is performed by six classifiers: HH, SS, TT, HS, ST and TH. Generally, the prediction of secondary structure is a three class (H, E, C) pattern recognition problem. The SVM method proposed in Gubbi et. al. [13] uses six classifiers which include three one vs one classifiers (H/E, E/C, C/H) and three one vs rest classifiers (H/~H, E/~E, C/~C). Multi-class classification is performed by combining the outputs of the six binary classifiers. Each of the six classifiers will read the data vector from the initial phase and generate corresponding classification result. Finally, the prediction result will be based on all of these six classification results in the final phase.

## IV. ARCHITECTURE AND DESIGN

This section presents the architecture of Jeeva, including the design of a web portal over the Aneka platform and its support for an SVM based prediction algorithm. We will briefly discuss background Aneka technology and its task programming model whose services are utilized in the realization of Jeeva portal.

### A. Aneka and Task Model

Aneka is a .NET-based enterprise Grid software platform, which allows the creation of enterprise Grid environments. Each Aneka node consists of a configurable container hosting several mandatory services and other optional services. The mandatory services provide the basic capabilities required in a distributed system, such as communications between Aneka nodes, security, and membership. Optional services can be installed to support the implementation of different programming models in Grid environments. For most programming models in Grid environments, their runtime system consists of a scheduler and many executors across distributed resources. For each model, its scheduler and executor are implemented as optional services in an Aneka container.

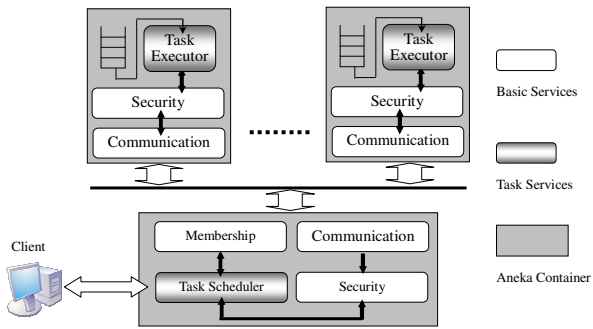

Fig. 2 Architecture of Aneka with Task Components.

Currently, Aneka supports the following programming models: thread model, task model, and MPI model. Thread and task models are used for independent tasks. In Jeeva, we choose task model to support the SVM-based algorithm.

Fig. 2 illustrates a configuration of Aneka deployment scenario for executing the task model. This is a representative setting of Aneka. One node is configured with a Task Scheduler component, while the other nodes are configured with Task Executor components. Basic service components, such as communication and security components are installed with every Aneka node for handling secure communications between them. A Membership service is typically hosted on the same Aneka node with the Scheduler component, which can query the Membership component for available Aneka nodes with Task Executor components.

By using this programming model, we can easily parallelize the SVM-based algorithm. A task is a single unit of work processed in a node, and is independent of other tasks executed on the same or on the other nodes at the same time. It is atomic, in the sense that it either executes successfully or fails.

During execution, a task (including its dependency for execution) is represented by an object, which can be serialized and submitted by the client to the scheduler. The task scheduler is always waiting for request messages such as task submission, query, and abort. Once a task submission is received by the scheduler, it is first queued and the scheduler thread picks up the queued tasks and maps them to available resources based on the configurable scheduling policy. Furthermore, the task scheduler keeps track of the queued and running tasks.

The task executor waits for task assignments from the scheduler. When the executor receives a task, it first unpacks the task object and its dependencies, creates a separate security context for the task, and then starts running the task. Once the execution of a task is finished, the executor sends the results back to the scheduler.

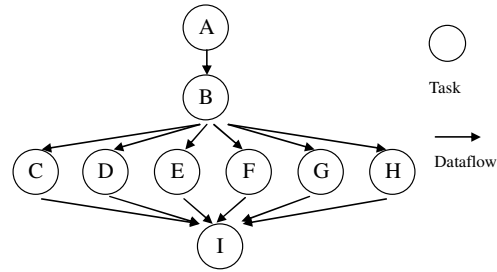

Fig. 3 Task Graph for SVM-based Algorithm.

To support the SVM-based algorithm in a parallel manner, we first subdivide the prediction process into multiple interdependent tasks. Fig. 3 shows the DAG (Directed Acyclic Graph) representation of the SVM-based algorithm. BLAST and Create Vector in the initial phase are represented by task *A* and *B* respectively. Tasks *C* to *H* represent 6 classifiers in the classification phase, while task *I* represents the final prediction phase. For each prediction job, the task client sends tasks from *A* to *I* to the task scheduler according to their dependency order. Within one job, tasks from *C* to *H* are totally independent and can be executed at the same time on different Aneka nodes. Furthermore, as the web portal is publicly shared, it may receive many prediction requests at the same time. For different requests, each task in one job is independent of the tasks in another job and they can be executed simultaneously.

*B. Design of Web Portal*

With the support of Aneka and its task model, we implemented task graph shown in Fig. 3 and developed a Web access interface. As illustrated in Fig. 4, our web portal system consists of two layers; namely web server layer and Aneka Grid layer. The web server layer is responsible for a) accepting protein secondary structure prediction requests from users; b) submitting prediction requests to Aneka Enterprise Grid for prediction and collecting prediction results; c) acknowledging prediction results to users, keeping prediction results in the database, and supporting online visualization in response to the queries of users. Aneka Grid layer supports its computing resources for prediction by means of a scalable and fault tolerant scheduling mechanism.

In the web server layer, we have one server machine which hosts an IIS (Internet Information Services) to provide portal services and an instance of task client for submitting task requests to Aneka Grid. Both input sets and the results need to be maintained in persistent storage so that users can retrieve results at later time. We have achieved this by recording all transactions in the database.

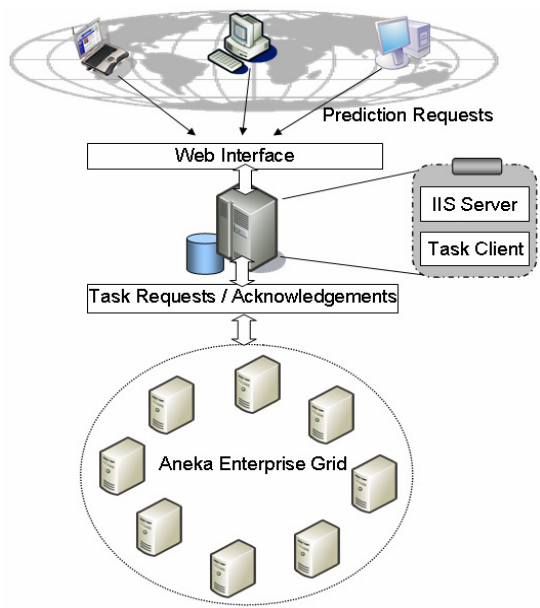

Fig. 4 Architecture of Jeeva.

The web portal accepts prediction requests from both anonymous and registered users. We provide an authentication service for registered users and keep the privacy of their results. For both anonymous and registered users, we keep their requests and results persistently in the database and provide a query service so that they can access their results online at any time. Additionally, the portal service also provides a management interface for the administrators, through which they can monitor the Aneka system and manage the information of users and prediction results in the database.

The task client in the web server layer works as a bridge between prediction requests and the Aneka computing services. The web interface first puts every prediction request into the database, and the task client frequently checks the database for new requests. Every time a new request is found, the task client generates a new job for the request and submits its tasks to Aneka according to the precedence order. For the task whose dependency consists of a large data set with infrequent changes, such as BLAST with the *nr* database which require about 2GB disk space, we deploy it on each Aneka node prior to its execution. During task submission, rather than sending the task with its large set of dependency to the task scheduler every time, we just send a request to execute BLAST. Similarly, what the executor receives from the scheduler is also an execution request, through which the executor invokes BLAST to execute locally. For other tasks, which may have frequent changes with small size of input data and dependency, such as each classifier, we serialize its content with its dependency modules and input data into one package and send it to the task scheduler.

The Aneka scheduler accepts task submissions and then maps them to the available Aneka nodes featuring the Task Executor component through a load balancing policy. Currently, the scheduler adopts a retry policy to handle failures. If one task fails due to physical machine failures, it will be rescheduled to other Aneka nodes. This process repeats until the task execution is completed successfully. Please refer to [27] for load balancing and failure handling policies in details.

*C. Implementation of Web Portal*

The web portal of Jeeva is implemented over ASP.NET platform and the task client is implemented with C# language over .NET framework.

Fig. 5 presents the interface for registered users to submit prediction requests. The prediction results are sent to the users through email. Furthermore, users can also browse their prediction history online. Fig. 6 illustrates one example prediction result through online browsing.

Fig. 5 Submit Prediction Request.

Detailed records of users and prediction results are stored in a SQL server. To enable easy discovery of bugs during the development, we keep a log for recording the error information of each task for every prediction job. The log is a text file in the file system of web server layer.

The administrators can monitor the status of the Aneka system with an Aneka web console, including the configuration of each Aneka node and the runtime performance statistics. As illustrated in Fig. 7, the detailed information of each machine is displayed when the mouse pointer moves over the icon.

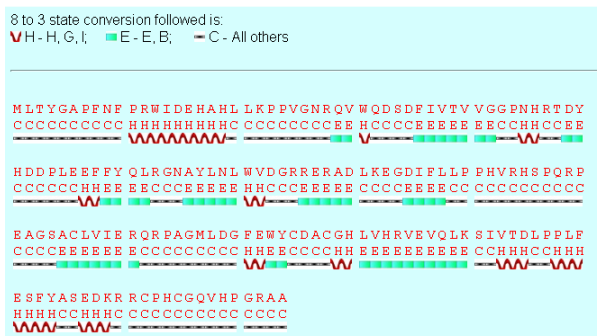

Fig. 6 Prediction Result.

Fig. 8 illustrates the performance statistics panel in the Aneka web console. The top panel displays the aggregated resource usage in the system, while the bottom panel displays the statistics on the tasks queues, including the waiting queue, running queue and finish queue.

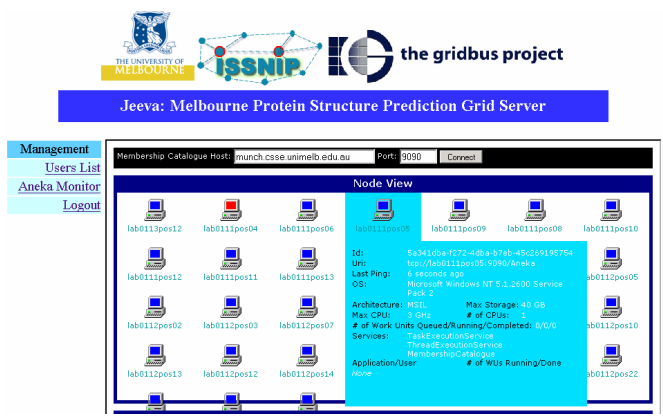

Fig. 7 System Monitor of Aneka Web Console.

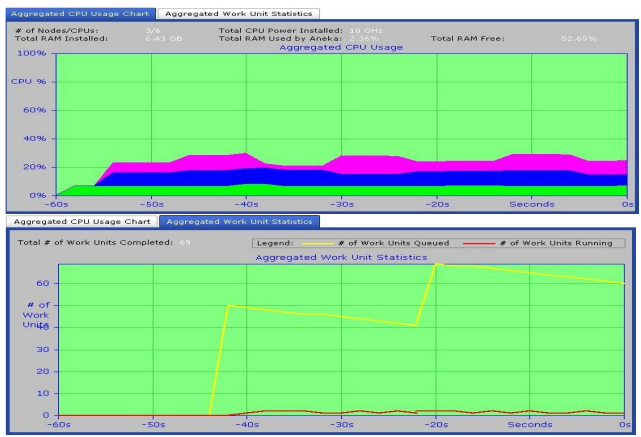

Fig. 8 Performance statistic of Aneka System.

The Aneka web console is implemented with Ajax. Every time when there are updates of the system status, an event is transferred through Ajax to the web console which displays the updated system status.

## V. PERFORMANCE EVALUATION

This section evaluates the performance of the backend runtime system of Jeeva. The experiments show the speedup of the SVM-based prediction algorithm deployed in Jeeva for single prediction job and the scalability of Jeeva system under multiple jobs submission. During the experiments, the Aneka system with task model for the protein secondary structure prediction was set in an enterprise Grid consisting of 37 nodes drawn from three student laboratories in the University of Melbourne. During testing, one machine worked in the web server layer hosting an IIS server and a task client. Other machines comprised Aneka system with one as a scheduler and the others as executors. Each machine has a single Pentium 4 processor, 500MB of memory, 160GB IDE disk, 1 Gbps Ethernet and runs Windows XP.

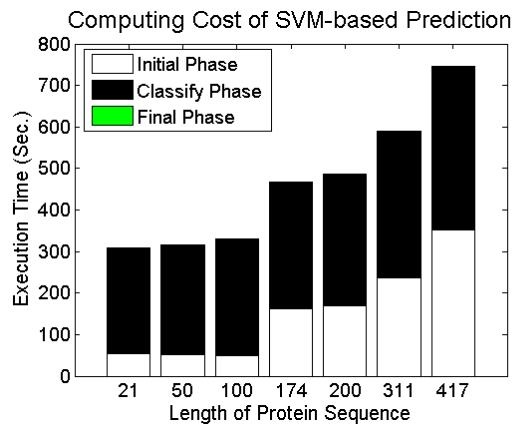

Fig. 9 Prediction Cost on Protein Sequences.

We conducted the experiments with the SVM-based prediction algorithm on the EVA dataset. The result gives an average $Q_3$ accuracy of 74.5% and ranks in top five protein structure prediction methods [13].

First, let us show the importance of parallelizing the classification phase for the SVM-based algorithm. Fig. 9 illustrates the performance of three phases of the SVM-based prediction for 7 protein sequences with different lengths. We can see that the time consumed by the classification phase dominates the time of whole prediction; the classification phase consumes 52.9% to 82.5% of the time of the whole SVM-based prediction. This phenomenon is more serious for protein sequences with a small length. Hence it is necessary to improve the efficiency of the classification phase.

We executed the parallelized SVM-based prediction algorithm for 4 protein sequences through the task model in Aneka with different numbers of executors. Fig. 10 illustrates the performance speedup. In the experiment, the lengths of 4 sample protein sequences are respectively 50, 100, 174 and 417. From the figure it is evident that the classification phase, which dominates the sequential execution time, decreases in the parallel version as the number of executors increases. With six Aneka executors, the execution time of the whole prediction algorithm is reduced by 65%~42%.

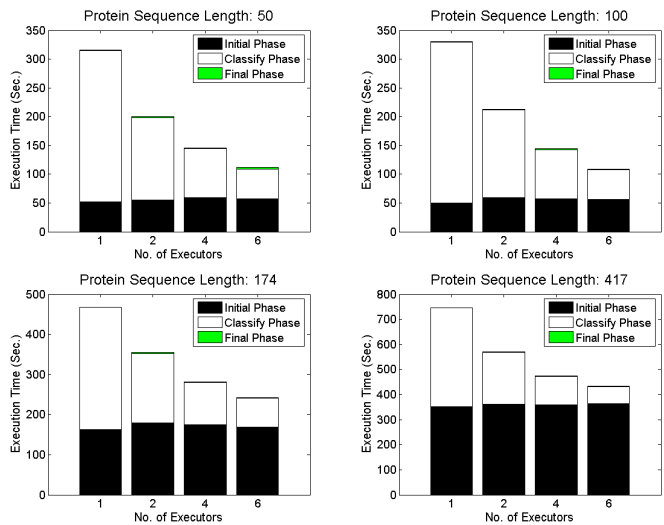

Fig. 10 Computing Cost of Parallel SVM-based Prediction Algorithm.

In the scalability experiment, we used 64 sample protein sequences. All of the 64 sequences were sent to the task client. After the task client received each sequence of prediction request, it created one job for it. Eventually there were 64 jobs created and sent to the Aneka scheduler. As illustrated in Fig. 11, the backend computing system of Jeeva is scalable with respect to the number of executors. Through 36 executors, the prediction on 64 samples was finished within 20 minutes.

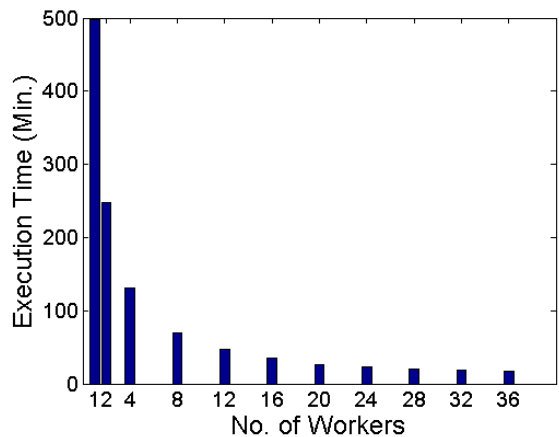

Fig. 11: Application execution time.

This section presents the architecture of Jeeva, including the design of a web portal over the Aneka platform and its support for an SVM based prediction algorithm. We will briefly discuss background Aneka technology and its task programming model whose services are utilized in the realization of Jeeva portal.

## VI. SUMMARY AND CONCLUSIONS

This paper presents Jeeva, a web portal for the protein secondary structure prediction, which is enabled by the Aneka platform. With the support of Aneka, an SVM-based prediction algorithm has been deployed in a parallel manner. The portal of Jeeva provides a convenient and flexible interface for both registered and anonymous users. Furthermore, administrators can also manage the history of prediction results through the web portal and monitor the running status of the Aneka system. The experiments were conducted to evaluate the speedup of the prediction algorithm and the scalability of Jeeva. We are working towards making the Jeeva portal for regular community use.

## ACKNOWLEDGMENT

This work is partially supported by research grants from the Australian Research Council (ARC) and Australian Department of Innovation, Industry, Science and Research (DIISR). We thank Christian Vecchiola, Mustafizur Ranhman, and Charity Laplap for their comments on improving the quality of the paper. We thank Alan Yim for his contribution to the development of Aneka monitoring web console.